\documentclass[prl,showpacs,floatfix,twocolumn]{revtex4}

\usepackage{graphicx}

\begin{document}

\title{Layering at liquid metal surfaces and interfaces:\\ Friedel oscillations and
confinement effects}

\author{Brent G.~Walker}
\affiliation{Condensed Matter Section,
Abdus Salam International Centre for Theoretical Physics,
Strada Costiera 11,
I-34014,
Trieste,
Italy.}

\author{Nicola Marzari}
\affiliation{Department of Materials Science and Engineering,\\
\mbox{Massachusetts Institute of Technology,
Cambridge,
MA 02139,
USA.}
}

\author{Carla Molteni}
\affiliation{Physics Department, King's College London,
Strand, London, WC2R 2LS, UK.}

\begin{abstract}

The structures of the liquid surface and the liquid-solid interface of
sodium have been characterized with extensive first-principles
molecular dynamics simulations. Friedel oscillations in the electronic
charge density at the free surface were found to persist across the
solid-to-liquid melting transition, with a small but distinctive
electronic layering that remains decoupled from the atomic
positions. Strong ionic layering was observed both at the liquid
surface and at the liquid-solid interface, notwithstanding the absence
of Friedel oscillations or under-coordinated atoms in the latter
case. Confinement effects at these soft or hard boundaries drive the
atoms into quasi-close-packed layers; even for this prototypical
free-electron metal Friedel oscillations are not relevant to ordering.

\end{abstract}

\pacs{61.25Mv,68.03.Hj,71.15.Mb,71.15.Pd}
\maketitle

Intensive theoretical
\cite{rice,iarlori,ditolla,celestini,fabricius,chacon,gonzalez} and
experimental \cite{expt1,expt10,expt11} work has focused on liquid
metal surfaces, showing the existence of \textit{surface-induced
layering}, where liquid atoms near the surface arrange themselves into
distinct layers. Such behavior is not present at the free surfaces of
dielectric or ionic liquids \cite{non-metal-surfaces}, while it is
observed for metallic and dielectric liquids confined by hard walls
\cite{rowlinson-widom}. These differences in the surface structure
mirror fundamental differences in bonding and screening. The interplay
of comparable energetic and entropic effects can then lead to a varied
thermodynamical phenomenology for the structure of liquid surfaces and
interface boundaries, including in-plane and out-of plane ordering
transitions, and reminiscent of the complex stability and phase
diagrams of crystalline surfaces.

Computer simulations have played a pivotal role in investigating the
nature and mechanisms of layering transitions at liquid metal
surfaces.  Effective-Hamiltonian simulations \cite{rice} suggested
that the rapid decay of the valence electronic density acts as a wall
against which atoms pack, in close analogy with the layer formation at
a solid-liquid interface \cite{rowlinson-widom}.  Glue-model
simulations of liquid Au \cite{iarlori}, Al and Pb \cite{ditolla} led
to the complementary suggestion that under-coordinated atoms at the
surface attempt to regain the favorable coordination of the bulk
liquid by moving inwards, increasing the density in the outermost
layer and causing a density oscillation to propagate into the
bulk. Density-functional theory (DFT) simulations of silicon (metallic
in its liquid phase) found instead that covalent-bond induced
correlations were responsible for the layering of the density
\cite{fabricius}. Recently it was suggested that metallic bonding does
not play an important role per se, and a layered density profile
should appear at the free liquid surface of any substance with a low
ratio of melting and critical temperatures (with the caveat that the
observation of layering is then usually preempted by solidification)
\cite{chacon}.  Last, it was proposed on the basis of first-principles
calculations for magnesium that electronic Friedel oscillations drive
the interlayer relaxations at solid metal surfaces \cite{cho}.

Prompted by these studies, we have decided to characterize the
structure of liquid surfaces and interfaces of sodium by means of
extensive first-principles calculations (over 250 ps in total, on
cells containing $\sim$ 160 atoms), using DFT paired with an accurate
generalized-gradient approximation to the exchange-correlation
potential. Sodium was chosen for its paradigmatic behaviour as a
nearly-free-electron metal, and for the challenges it poses to
experiments due to its low surface tension that leads to
thermally-excited capillary waves. While recent results have appeared
for Na:K alloys \cite{expt10} and pure K \cite{expt11}, layer
formation was extrapolated from the lead-up to a peak in the x-ray
reflectivity, rather than the direct observation of a full peak, as
was the case for all other metals for which layering was conclusively
observed (Ga, In, Hg) \cite{expt1}. The liquid surfaces of Na, Li
\cite{gonzalez}, and Na:K alloys \cite{gonzalez2} have also been
recently studied from first-principles -- albeit using approximate
orbital-free versions of DFT based on kinetic energy functionals
\cite{madden1} -- and ionic layering at the surface was observed.

In our calculations, we used state-of-the-art approaches
\cite{nicolapaper} to first-principles molecular dynamics (MD) to
examine the microscopic structure and dynamics at the free surface or
interface. More importantly, we devised a series of {\it gedanken}
computer experiments to elucidate the role and nature of the
electronic and ionic responses to a surface or interface.

First, we performed extensive simulations of the Na liquid surface at
two temperatures ($400$~K and $500$~K) above the melting point
($371$~K).  We started with two bulk structures corresponding to a bcc
crystal containing either 160 atoms arranged in 10 layers along the
[001] direction or 162 atoms in 18 layers along the [111]
direction. We chose two different geometries to rule out any influence
of the supercell cross-section on the intra-layer atomic
ordering. These systems were melted and equilibrated at high
temperature; then, a vacuum layer of thickness $\sim 11$~\r{A} was
inserted along the the [001] or [111] direction to create two liquid
slabs from the bulk samples. MD simulations for each of these two
slabs were performed for 50 ps (20 ps of equilibration and 30 ps of
data collection) at 400 K and 500 K \cite{na_paper_large}. Careful
consideration was given to finite-size effects, by (a).~analyzing the
density-density autocorrelation function \cite{fabricius}, which
demonstrated independence of the oscillations propagating from the two
opposite surfaces in each slab, and (b).~running a series of classical
MD simulations on increasingly larger systems, which established that
the systems chosen were large enough to avoid any effect of one
surface on the local structure of the other surface, and small enough
not to have layering washed away by capillary waves
\cite{brentthesis}. All simulations used ensemble-density functional
theory \cite{nicolapaper} and a 0.5 eV cold-smearing generalized
entropy \cite{nicolapaper1999}, a formulation particularly efficient
for metals. We used Troullier-Martins pseudopotentials
\cite{troullier-martins}, the PW91 generalized-gradient approximation
for the exchange and correlation functional, non-linear
core-corrections and the Baldereschi point for Brillouin zone
integrations \cite{pw91_nlcc_baldereschi}.

\begin{figure}[!hbt]
{\centering
\includegraphics[clip,width=\columnwidth]{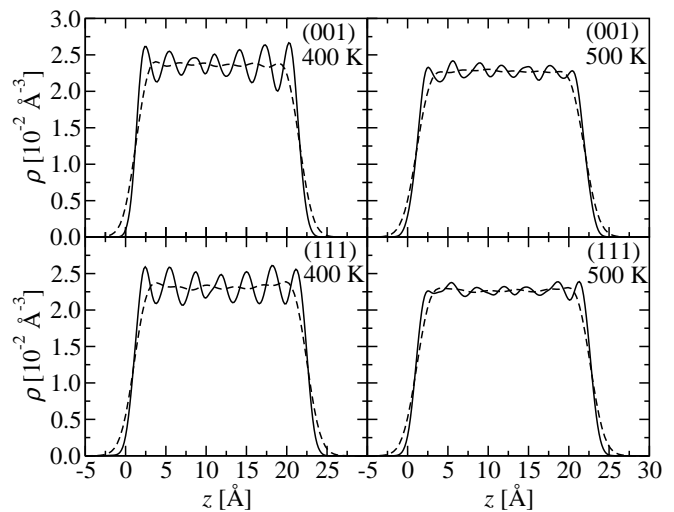}
\caption{\label{density_profiles_figure} Atomic (solid) and electronic
  (valence only, dashed) density profiles for the different liquid
  slabs, plotted normal to the liquid surfaces for simulations
  performed at 400 K and 500 K. (001) and (111) denote the
  cross-section of the supercell (see text).} }
\end{figure}

Fig.~\ref{density_profiles_figure} shows the density profiles normal
to the liquid slab; clear oscillations in the atomic densities are
visible, indicating layer formation. Oscillations in the valence
electronic densities are also seen, corresponding roughly to the
interlayer regions. Layering is clearly evident in the simulations at
400 K; above that temperature, entropy starts to overcome the ordering
in the liquid, and at 500 K the ionic peaks are already significantly
reduced. The atomic density profiles indicate formation of 7 layers
parallel to the surface; while the finite thickness of our slabs meant
that layers appeared throughout the systems, we stress again that
classical simulations showed that the structure and layering at the
surface would not change if more layers were introduced.

Second, in order to establish whether layering is driven by electronic
screening, confinement, or densification following decreased
coordination at the surface, we examined -- in a computational proof
of principle -- two different case studies for which these effects can
be decoupled.

\begin{figure}[!bt]
\begin{center}
\includegraphics[clip,width=\columnwidth]{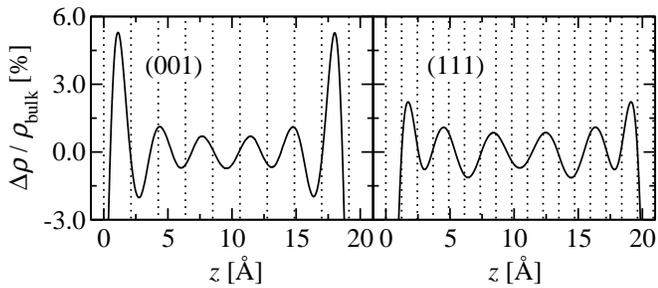}
\end{center}
\caption{\label{charge_redist_fig2} Electronic (valence) charge
density response to the creation of surfaces in a perfect crystalline
sample of Na, cleaved to expose either the (001) or (111)
surfaces. The dashed lines indicate the positions of the atomic layers
(10 layers in the (001) supercell and 18 in the (111) supercell).}
\end{figure}

We first monitored the electronic response of a perfect solid and of a
disordered bulk liquid to the creation of surfaces. These surfaces
were obtained, for any given atomic configuration, by cleaving a
sample in half and rigidly separating the two fragments with the
insertion of a large amount of vacuum. The atoms were not allowed to
relax, and thus all the screening comes from the reorganization of the
electronic charge density in response to the appearance of surface
boundaries. We calculated the charge redistribution induced by such
perturbations: electronic densities for the bulk and the cleaved
systems were first individually determined, and then subtracted to
obtain direct information on the electronic screening \cite{cho}.  In
Fig.~\ref{charge_redist_fig2} we show the results for the case of an
ordered solid, where, starting from a bcc Na crystal, characteristic
Friedel oscillations appear following the creation of ideal (001) or
(111) surfaces. These results are in close agreement with the findings
for Mg of Ref.~\cite{cho}, and highlight the remarkable appearance of
Friedel oscillations due to the response of the valence electrons to
the perturbation created by the surface boundary \cite{a&m}.

\begin{figure}[!htb]
{\centering
\includegraphics[clip,width=\columnwidth]{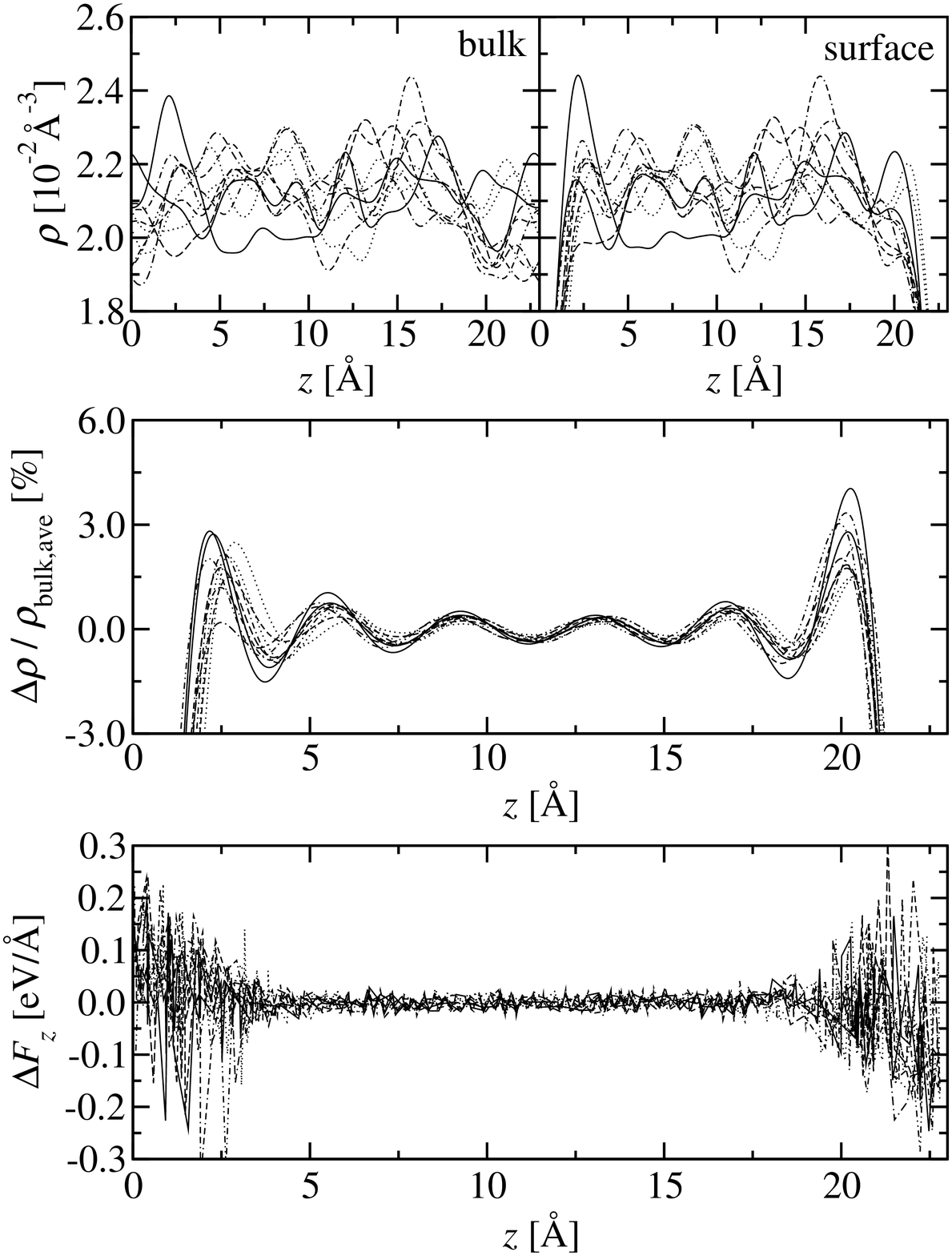}

}
\caption{\label{charge_redist_MD} Top panel: electronic (valence)
charge densities for uncorrelated configurations taken from a bulk
liquid MD simulation; the bulk densities are on the left, and the
corresponding cleaved ones are on the right. Middle panel: charge
density differences between the bulk and cleaved samples, showing
clearly the appearance of Friedel oscillations. Bottom panel:
differences between the bulk and cleaved samples in the
Hellmann-Feynman forces acting on the ions in the direction
perpendicular to the cleaved surfaces.}
\end{figure}

Next, we performed the same analysis for thermally-equilibrated bulk
liquid samples. We took atomic configurations from a number of
uncorrelated timesteps of a 500 K MD simulation of the bulk liquid,
and for each configuration inserted a thick vacuum layer. Again, the
atomic positions were kept frozen, so that no relaxation or
thermalization of the atoms was allowed and thus no atomic layers were
present. The valence electronic densities for these configurations are
shown in the top panels of Fig.~\ref{charge_redist_MD} for a set of
bulk configurations and their cleaved counterparts. Remarkably, when
charge-density differences are taken, very clean and distinct Friedel
patterns are recovered (Fig.~\ref{charge_redist_MD}, middle panel)
even for such disordered, ``bulk liquid''-like surfaces, in close
analogy with the cleaving of perfectly-ordered solids. This result
proves the persistence of electronic Friedel oscillations through the
melting and disordering of the solid. It also underlines how, for this
simple metal, the effect of the atomic positions on the electronic
response is minimal. To confirm this last point we examined the
differences in the Hellmann-Feynman forces (in the direction
perpendicular to the surfaces) for Na atoms in the bulk-liquid and
cleaved configurations. These differences are shown in the bottom
panel of Fig.~\ref{charge_redist_MD}, and highlight the absence of
Friedel-like patterns in the atomic forces. While differences in the
atomic forces in the region where surfaces were created are clearly
visible, these effects decay rapidly, and the atoms away from the
surface by more than 2-3 \r{A} would not be significantly affected or
modulated in their dynamics by the electronic response.

\begin{figure}[!b]
{\centering
\includegraphics[clip,width=\columnwidth]{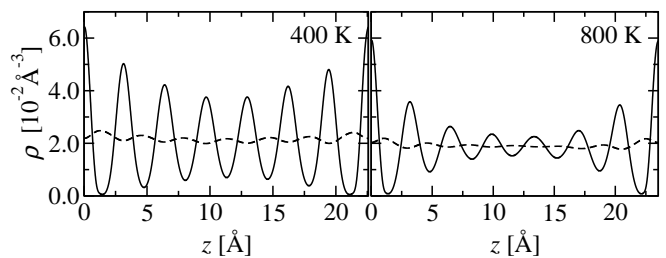}
\caption{\label{confine_figure} Atomic (solid) and electronic (valence
only, dashed) density profiles for the confined liquid. The (001)
fixed monolayers are at $z=0$~\r{A} in both simulation cells.}}
\end{figure}

In the second part of our proof of principle we considered instead a
confined bulk liquid. We created in our bulk simulations (using the
(001) supercell geometry of 160 atoms described earlier) a fixed layer
of sodium atoms: given that periodic boundary conditions were in
place, the bulk liquid metal was then confined between two infinite
2-dimensional hard walls. Since these walls were only one-layer thick,
and composed of metallic sodium at a density practically identical to
that of the liquid, they would not give rise to any charge density
discontinuity and associated Friedel oscillations. Also, no loss of
coordination was experienced by the atoms in the liquid close to the
interface, and thus there was no driving force towards densification
at the liquid boundary. The fixed layers were created by rearranging
24 of the 160 atoms in the bulk liquid supercell into an hexagonal
planar structure (slightly strained in one direction, to accommodate
the square cross-section of the (001) simulation cell). MD simulations
were then performed at 400~K and 800~K. We show in
Fig.~\ref{confine_figure} the atomic and valence electronic densities
obtained.
Again, the distinctive formation of layers is observed, persistent now
even at the high temperature of 800 K. In perfect analogy with the
case of the liquid slabs, the atoms here rearranged themselves into 7
distinct layers. Thus, we have again obtained layering for the sodium
atoms close to a boundary, notwithstanding the absence of Friedel
oscillations. This consideration, together with the previously
observed lack of modulation in the ionic forces attributable to the
electronic response, rules out Friedel oscillations in driving ionic
layering. In addition, in the confined liquid sample -- as opposed to
the free surface -- there were no under-coordinated atoms and no
driving force to increase the density of the surface and to initiate
the formation of a distinctive surface layer. Thus, a liquid analogue
of the densification observed at the surface of open solid surfaces
and the associated interlayer contraction \cite{heinefinnis} also
needs to be ruled out, pointing finally at geometric confinement
effects as the foremost and common cause of layering oscillations both
at the liquid surface and at the liquid-solid interface of simple
metals. These simple geometric considerations are reinforced by the
study of the pair distribution functions in the surface plane, that
show a nearest-neighbor distance of $\sim 3.8$~\r{A}, corresponding to
24-25 atoms in a quasi-hexagonal arrangement (the layers are not
perfectly packed, with in-plane coordination numbers averaging between
4.5 and 5.3). Thus, for systems containing 160-162 atoms, we expect a
number of close-packed layers between 6 and 7, consistent with what
was found for the free and confined liquid systems. Close-packed
layers at this density would be spaced by $\sqrt{2/3} \times
3.8~\text{\r{A}}= 3.1~\text{\r{A}}$, which compares favorably with our
observed interlayer spacings of $2.98 \pm 0.23$~\r{A} and $3.02 \pm
0.25$~\r{A} in the liquid at 400 K and 500 K respectively, and $3.23
\pm 0.10$~\r{A} and $3.36 \pm 0.23$ for the liquid-solid interface at
400 K and 800 K.
 
In summary, we have characterized with extensive first-principles MD
simulations the free liquid surface and the liquid-solid interface of
sodium, chosen as a paradigmatic example of a free-electron metal. We
found clear signatures of Friedel oscillations at the solid surfaces
that remarkably persist through the transition to a liquid
surface. These oscillations are decoupled from the underlying atomic
positions, and do not appreciably affect the ionic forces on the
system. Very clear ionic layering was observed for the liquid
surfaces, and an even stronger layering was found for the liquid
confined by its own solid. Given that Friedel oscillations are absent
in the latter case, and that a homogeneous valence charge density
rules out any driving force for densification at the boundary, we
conclude that geometric confinement is the foremost cause of
layering. The confinement induced by the presence of a solid boundary
is stronger than in the case of a vacuum boundary, and layering is
correspondingly stronger for the liquid bound by its own solid.

B.~G.~W.~thanks the New Zealand Foundation for Research Science and
Technology and the Cambridge Overseas Trust for research studentships.

\bibliographystyle{apsrev} \bibliography{na_paper_prl}

\end{document}